School neighbourhood and compliance with WHO-recommended annual NO$_2$ guideline: a case study of Greater London

Niloofar Shoari[a*], Shahram Heydari[b], Marta Blangiardo[a]

[a] *MRC Centre for Environment & Health, Department of Epidemiology and Biostatistics, Imperial College London, London, UK*

[b] *Department of Civil, Maritime, and Environmental Engineering, University of Southampton, Southampton, UK*

**Abstract**

Despite several national and local policies towards cleaner air in England, many schools in London breach the WHO-recommended concentrations of air pollutants such as NO$_2$ and PM$_{2.5}$. This is while, previous studies highlight significant adverse health effects of air pollutants on children's health. In this paper we adopted a Bayesian spatial hierarchical model to investigate factors that affect the odds of schools exceeding the WHO-recommended concentration of NO$_2$ (i.e., 40 µg/m$^3$ annual mean) in Greater London (UK). We considered a host of variables including schools' characteristics as well as their neighbourhoods' attributes from household, socioeconomic, transport-related, land use, built and natural environment characteristics perspectives. The results indicated that transport-related factors including the number of traffic lights and bus stops in the immediate vicinity of schools, and borough-level bus fuel consumption are determinant factors that increase the likelihood of non-compliance with the WHO guideline. In contrast, distance from roads, river transport, and underground stations, vehicle speed (an indicator of traffic congestion), the proportion of borough-level green space, and the area of green space at schools reduce the likelihood of exceeding the WHO recommended concentration of NO$_2$. As a sensitivity analysis, we repeated our analysis under a hypothetical scenario in which the recommended concentration of NO$_2$ is 35 µg/m$^3$ – instead of 40 µg/m$^3$. Our results underscore the importance of adopting clean fuel technologies on buses, installing green barriers, and reducing motorised traffic around schools in reducing exposure to NO$_2$ concentrations in proximity to schools. This study would be useful for local authority decision making with the aim of improving air quality for school-aged children in urban settings.

1. Introduction

Air pollution is recognised as a serious environmental risk factor for child health and survival by the World Health Organization (WHO) [1]. Children are particularly vulnerable to high concentrations of air pollutants because of their rapidly growing organs and high rates of respiration [2, 3]. Scientific community has provided evidence on links between exposure to air pollution in early life and increased risk of developing or exacerbating respiratory symptoms including asthma, bronchitis, and pneumonia [4-9], reduced lung function [2], reduced cognitive ability [10], emerging mental health problems such as attention deficit/hyperactivity disorder, anxiety, and depression [11, 12], reduced rate of school absenteeism, and higher rates of failure in educational tests [13].

In the UK, one in three children are growing up in areas with elevated concentrations of air pollutants such as PM$_{2.5}$ and NO$_2$ [14]. Monitoring personal exposure of 250 pupils to NO$_2$ and PM$_{2.5}$ in London through wearable sensors, researchers have found that children are exposed to high pollution levels at schools, and even more when commuting from/to schools [15]. Breathing polluted air will have long-term health effect on individuals and populations. For example, unsafe air quality in London restricts school-age children lung growth (a loss of 5 per cent in lung capacity over a period of five-year), putting them at the risk of lung diseases in their adult life [16]. The negative impact of $NO_2$ concentrations on lung function growth in childhood has been reported in several cities of Europe, North America, and China [17, 18].

[*] Corresponding author's email: n.shoari@imperial.ac.uk

Transport is recognized as one of the major contributors to $NO_2$ levels [19]. Several interventions to tackle air quality issues are introduced in London with the aim of reducing motorised traffic in general as well as reducing primary $NO_2$ emitted from old diesel cars. For example, the Low-Emission Zone (LEZ) and Ultra-Low-Emission Zone (ULEZ) interventions ban all polluting vehicles to enter central London. Nevertheless, these interventions were only marginally helpful [20], and many schools remain in highly polluted areas and in proximity to heavily travelled roadways and commercial premises. Therefore, school microenvironments constitute a fundamental part of pupils' diurnal exposure to air pollutants.

Concentrations of air pollutants such as $NO_2$ can vary even among schools located in the same local authority due to variability in certain attributes of schools and their surroundings [21]. Existing literature highlights that schools close to roadways and located in neighbourhoods with a higher road density exhibit higher pollution levels, and that green space surrounding schools, usually measured by normalised difference vegetation index and tree canopy, provides protective effects against air pollution [22-28]. However, an integrative approach from socioeconomic, household, transport, land use, built and natural environment perspectives to assess the link between school neighbourhood attributes and air pollutant concentrations is missing. Understanding this link is important as it enables us to identify the most important contributing or protective factors that affect concentrations of $NO_2$ around schools. This is essential for devising local interventions to complement the existing strategies. In this paper we employed a Bayesian hierarchical (multilevel) binary logit model to quantify the magnitude of the effects of various factors on the odds of exceeding the WHO-recommended annual concentration of 40 µg/m$^3$ for $NO_2$ at various spatial resolutions. The current WHO target concentration is not a standard below which there is no risk to public health but merely is a starting point for developing country-specific and long-term guidelines. We also discuss implications resulting from a hypothetical scenario where the guideline concentration for $NO_2$ is reduced to 35 µg/m$^3$. Finally, we provide evidence-based recommendations for local interventions to contain pupils air pollution exposure during school hours.

2. Data

We created a comprehensive data set containing several variables of different types obtained from various sources. The latest air pollution data, available for the year 2016, was provided by London Atmospheric Emissions Inventory (LAEI), containing annual average $NO_2$ and $PM_{2.5}$ concentrations represented by points over a 20m x 20m grid across Greater London [29]. School data for 2016 was obtained from the Ordnance Survey (OS) MasterMap Sites Layer product, which includes the locations and boundaries of all educational establishments' grounds. We restricted our analysis to schools with pupils aged from 5 to 16 years, excluding educational establishments functioning as higher education institutions (universities and colleges). This resulted in 2,861 schools in Greater London in 2016, covering around 1.4 million children.

We conducted a comprehensive desk-based data collection campaign to obtain a host of explanatory variables (socioeconomic, household, transport, land use, built and natural environment characteristics) at school- and area-level. Our area-level variables represented measures from electoral wards (as proxy for neighbourhood characteristics) and boroughs. Based on the latest ward boundary classification in 2014, there are 630 wards in Greater London with a mean area of 2.5 km$^2$ and average population of around 14,000 per ward.

Table 1 reports descriptive statistics of these variables together with their sources. Transport-related variables at school-level included the distance between school and the closest main road, ferry station, and underground station, intensity of total roads within buffers of 100m and 400m around schools, the number of traffic lights within buffers of 100m and 400m around schools, and the number of bus stops within buffers of 100m and 400m around schools. The area-level variables included average vehicle speed, annual underground passenger counts of entries and exits, annual average daily traffic flow, and transport fuel consumption.

We calculated the straight-line distance between each school polygon and the nearest main road (defined as A-road and motorways in OS Open Roads data), ferry station (provided by OS Points of Interest database), and underground station (from Transport for London). The intensity of roads within 100m and 400m buffer around each school was computed by dividing the sum of the all road segments, generated from OS Open Roads data, by the area of each buffer (i.e., total length per area). The location of traffic lights was obtained from the OpenstreetMap (OSM) database, which provides fairly complete and accurate geographical data in London as indicated by [30]. The input on the location of bus stops was extracted from OS Point of Interest layer by considering entries classified as "bus stop" under the features labelled with "bus transport category". We then calculated the total number of traffic lights and bus stops within 100m and 400m circular buffers around each school. These buffers were selected based on previous research [31-33].

We used geocoded information of underground stations, including passenger entry and exit statistics for 2016 available from London Underground Performance Reports by Transport for London (TfL). The LAEI 2013 version contains some traffic-related data including vehicle speed at major road segments that were derived from GPS-based vehicle tracking systems. We used these data to calculate the average vehicle operating speed in each ward. The data on annual average daily traffic flow for boroughs in 2016 were provided by the Department for Transport (DfT). We retrieved the 2016 data on road transport fuel (petrol and diesel) consumption from the Department for Business, Energy and Industrial Strategy.

To have an insight on the association between natural environment and local air pollution concentrations, we included the total area of green space available at both school and ward levels. To estimate green space available at London schools, we used high-resolution land use data from OS Topographic layer and calculated green space by isolating land features within school boundaries that are covered by vegetation. The details of the methodology is explained in [34]. For ward-level green space area, we used OS Open Green space data, which provides information on the location and extent of public parks, playing fields, sports facilities, play areas and allotments, and calculated the total area of green space in each ward. In terms of built environment variables, we included the population density of boroughs (as an indicator of human activity patterns) obtained from Greater London Authority (GLA) population estimates for 2016 and land use characteristics at borough-level from Generalized Land Use Database (GLUD). From the household and socioeconomic characteristics, we considered various variables such as the percent of children in poverty, the percent of lone parents not in employment, the type of household (e.g., detached house), and the type of tenure (e.g., owned), derived from the 2011 UK census.

### 2.1 Data Preparation

To estimate school air pollution levels, we aggregated concentration data over schools' boundaries and estimated average $NO_2$ and $PM_{2.5}$ concentrations for all schools. There were 22 cases of small schools where the pollution point did not cover the school boundary, for which the average concentrations within a 20m buffers around schools was used. We then dichotomized schools' pollution levels using the WHO-recommended concentration of 40 µg/m$^3$ for $NO_2$ and 10 µg/m$^3$ for $PM_{2.5}$. The annual average $NO_2$ concentrations at schools ranged from 23.79 µg/m$^3$ to 61.67 µg/m$^3$. Around 17% of London schools (equivalent to 495) exceeded the WHO-recommended concentration of $NO_2$. For $PM_{2.5}$, school-level annual concentrations ranged from 11.40 µg/m³ to 16.20 µg/m³, exceeding the recommended concentration in all London schools. Therefore, we only focus on $NO_2$ in this paper.

Table 1. Explanatory variables considered and their descriptive statistics for Greater London

| Variable | Spatial unit | Mean (SD) | Min | Max | Source |
|---|---|---|---|---|---|
| *Transport-related variables* | | | | | |
| Bus stops within 100m buffer | School | 1.54 (1.64) | 0.00 | 19.00 | OS-POI[a] |
| Bus stops within 400m buffer | School | 11.73 (5.79) | 0.00 | 53.00 | |
| Traffic lights within 100m buffer | School | 1.14 (2.99) | 0.00 | 27.00 | OSM[35] |
| Traffic lights within 400m buffer | School | 10.74 (15.96) | 0.00 | 130.00 | |
| Distance to main road (km) | School | 0.25 (0.28) | 0.01 | 2.78 | OS-OR[a] |
| Distance to ferry station (km) | School | 6.05 (4.32) | 0.05 | 19.07 | OS-POI[a] |
| Distance to underground station (km) | School | 2.59 (3.08) | 0.01 | 18.79 | TfL[36] |
| Intensity of roads within 100m buffer (m/m²) | School | 0.02 (0.01) | 0.00 | 0.03 | OS-OR[a] |
| Intensity of roads within 400m buffer (m/m²) | School | 0.02 (0.01) | 0.00 | 0.03 | |
| Average vehicle speed (kph) | Ward | 29.46 (5.42) | 15.10 | 54.56 | LAEI 2013 [37] |
| Bus fuel consumption (in 1000 tonnes of oil equivalent) | Borough | 4.69 (2.33) | 1.75 | 13.69 | |
| Diesel cars consumption (in 1000 tonnes of oil equivalent) | Borough | 21.27 (9.26) | 4.73 | 50.68 | |
| Petrol cars consumption (in 1000 tonnes of oil equivalent) | Borough | 26.66 (10.34) | 3.82 | 57.74 | |
| HGV consumption (in 1000 tonnes of oil equivalent) | Borough | 8.88 (7.68) | 1.75 | 35.60 | BEIS[b] |
| Diesel LGV consumption (in 1000 tonnes of oil equivalent) | Borough | 12.33 (4.78) | 2.55 | 25.47 | |
| Petrol LGV consumption (in 1000 tonnes of oil equivalent) | Borough | 0.53 (0.18) | 0.11 | 1.01 | |
| Motorcycles fuel consumption | Borough | 1.02 (0.40) | 0.36 | 2.30 | |
| Overall consumption (in 1000 tonnes of oil equivalent) | Borough | 75.4 (30.54) | 15.40 | 159.89 | |
| Annual average daily traffic flow | Borough | 6,139 (1,356.80) | 4,282 | 9,002 | DfT[c] |
| Underground annual entries & exits (million) | Borough | 106.67 (132.53) | 9.12 | 656.55 | TfL[36] |
| *Land use, built and natural environment variables* | | | | | |
| Area of greenspace at schools (m²) | School | 10.94 (16.96) | 0.00 | 242.05 | OS-Topo[a] |
| Area of green space at ward (ha) | Ward | 40.17 (63.99) | 0.00 | 605.15 | OS-GS[a] |
| Population density (per hectare) | Borough | 74.48 (39.38) | 21.84 | 155.62 | GLA[38] |
| Land use with green space (%) | Borough | 30.91 (12.72) | 4.83 | 59.32 | |
| Land use with domestic buildings (%) | Borough | 10.23 (3.31) | 5.01 | 19.2 | |
| Land use with non-domestic buildings (%) | Borough | 7.25 (6.58) | 1.63 | 37.50 | GLUD [39] |
| Land use with domestic gardens (%) | Borough | 22.37 (7.62) | 0.12 | 34.74 | |
| Land use with water (%) | Borough | 3.45 (4.87) | 0.10 | 22.20 | |
| *Household and socioeconomic variables* | | | | | |
| Detached house (%) | Ward | 6.57(0.08) | 0.28 | 55.83 | GLA [40] |
| Semi-detach house (%) | Ward | 19.82 (0.16) | 0.19 | 82.30 | |
| Terraced house (%) | Ward | 23.56 (0.13) | 1.45 | 63.83 | |
| Flat/apartment (%) | Ward | 50.05 (0.25) | 6.26 | 97.94 | |
| Number of households owned | Ward | 1095 (498) | 132 | 2903 | |
| Number of households owned with mortgage/loan | Ward | 1474 (502) | 345 | 3068 | |
| Number of households socially rented | Ward | 1248 (850) | 44 | 4214 | |
| Number of households privately rented | Ward | 1300 (640) | 191 | 3748 | |
| Number of households rent free | Ward | 68 (39) | 21 | 388 | |
| Mean annual household income (per £1,000) | Ward | 45.59 (15.58) | 19,27 | 143.81 | |
| Children in poverty (%) | Borough | 19.27 (6.02) | 8.8 | 32.5 | GLA[38] |
| Lone parents without employment (%) | Borough | 46.11 (8.54) | 20.82 | 73.58 | |

[a] Ordnance Survey Point of Interest (OS-POI), OS Open Green space (OS-GS), OS Open Roads (OS-OR), OS Topography layer (OS-Topo) data are available from https://digimap.edina.ac.uk/
[b] Department for Business, Energy and Industrial Strategy (BEIS) available from https://www.gov.uk/government/collections/sub-national-consumption-of-other-fuels
[c] https://www.gov.uk/government/statistical-data-sets/road-traffic-statistics-tra#traffic-volume-in-kilometres-tra02

### 3. Method

In this paper we are concerned about the odds of schools exceeding specific threshold values of $NO_2$ concentrations. Therefore, a binary logit approach is suitable. We model this data using a Bayesian hierarchical logistic regression model. Our model accounts for spatial dependence at ward level as well as for the nested structure of the data (schools nested within wards).

Let $Y_{sw}$ be a binary outcome indicating whether a school $s$ (s=1,2,...,2861) in ward $w$ (w=1,2,...,630) is non-compliant with the threshold value (e.g., the WHO-recommended concentration of 40 µg/m³). We can then write

$$Y_{sw}|p_{sw} \sim Bernoulli\ (p_{sw}) \\ logit(p_{sw}) = \alpha + \boldsymbol{X_{sw}\beta} + \boldsymbol{Z_w\gamma} + \boldsymbol{T_b\theta} + U_w + V_w + e_s \quad (1)$$

where $p_{sw}$ represents the odds of exceeding the threshold value of interest for NO₂, α is the global intercept; $\boldsymbol{X_{sw}}$ denote the vector of school-level attributes with their respective coefficients $\boldsymbol{\beta}$; $\boldsymbol{Z_w}$ and $\boldsymbol{T_b}$ denote, respectively, the vectors of ward- and borough-level attributes with their respective coefficients $\boldsymbol{\gamma}$ and $\boldsymbol{\theta}$. To account for residual spatial variability, we considered a Besag, York and Mollié (BYM) specification [41] that includes two random effects $V_w$ and $U_w$. $V_w$ is an unstructured random effect at ward-level, $V_w \sim Normal(0, \tau_V^{-1})$; and $U_w$ denotes a structured ward-level random effect that allows for spatial dependency between neighbouring wards. For $U_w$ we considered an intrinsic conditionally autoregressive (ICAR) structure as follows:

$$U_w|U_{-w} \sim Normal(\frac{\sum_k a_{ik} U_k}{N_w}, (N_w \tau_U)^{-1}) \quad (2)$$

where $U_{-w}$ denotes the rest of the wards excluding the $w$th wards. This term implies that spatial error in ward $w$ follows a normal distribution with a mean equal to the average error of neighbouring wards ($U_k$), where $a_{ik}$ is 1 if the k-th ward is a neighbour and 0 otherwise; the variance $\tau_U$ is inversely proportional to the number of a ward neighbours ($N_w$). The term $e_s$, accounts for extra variability at school level and follows a normal distribution with the mean 0 and the variance $\tau_s$.

### 3.1 Prior specification and model computation

We specified non informative Normal (0,100) prior distributions for each element in $\boldsymbol{\beta}$, $\boldsymbol{\gamma}$, and $\boldsymbol{\theta}$, Gamma(1, 0.001) for the inverse of the variance $\tau_s$, Gamma(0.1, 0.1) for the inverse of the variance $\tau_V$, and Gamma(0.5,0.005) for the inverse of spatially structured variance of $U$ – as it is common in the specification of conditionally autoregressive models. Inference was performed through Markov Chain Monte Carlo (MCMC) simulations in the *NIMBLE* Package in R [42]. We centred covariates at their observed mean to facilitate the convergence of our MCMC simulations. The posterior inferences for model parameters were obtained from two chains with 100,000 iterations and the first 30,000 iterations were discarded to ensure convergence. We checked the convergence of the parameters using the Gelman-Rubin statistic and visually using history plots.

### 4. Results

We provide the results focusing on the estimates of odds ratios and the estimated expected probabilities of exceeding the WHO-recommended concentration of NO₂ at both ward and borough (local authority) levels. We chose variables based on the extant literature and a backward stepwise variable selection. We tested several sets of variables, avoiding highly correlated ones (more than 60%) in the model at the same time. Table 2 reports the regression coefficient estimates, indicating that several variables of different types have a bearing on the likelihood of schools exceeding the WHO-recommended concentration of NO₂.

Table 2. Posterior estimates of regression coefficients

|  | Mean (SD) | 95% credible interval |
|---|---|---|
| *Transport-related variables* |  |  |
| Bus stops within 100m buffer | 0.33 (0.08) | 0.18, 0.49 |
| Traffic lights within 400m buffer | 0.04 (0.01) | 0.02, 0.06 |
| Distance to main road (km) | -0.46 (0.10) | -0.66, -0.27 |
| Distance to ferry station (km) | -0.36 (0.10) | -0.56, -0.17 |
| Distance to underground station (km) | -0.67 (0.20) | -1.07, -0.30 |
| Average vehicle speed (kph) | -1.22 (0.27) | -1,78, -0.71 |
| Bus fuel consumption | 0.32 (0.09) | 0.15, 0.52 |
|  |  |  |
| *Land use, built and natural environment variables* |  |  |
| Area of green space at schools | -0.25 (0.07) | -0.40, -0.12 |
| Percent land use with greenspace | -0.08 (0.03) | -0.14, -0.03 |
|  |  |  |
| *Household and socioeconomic variables* |  |  |
| Percent Lone parents without employment | 0.09 (0.03) | 0.03, 0.14 |
| Percent terraced houses | -0.12 (0.02) | -0.17, -0.08 |
|  |  |  |
| Variance of school-level error term | 0.02 (0.01) | 0.01, 0.23 |
| Variance of ward-level unstructured random effects | 3.62 (0.06) | 1.79, 6.32 |
| Variance of ward-level structured random effects | 0.21 (0.50) | 0.01, 2.04 |

### 4.1 Estimates of odds ratios

For practical interpretation of the regression coefficients reported in Table 2, we calculated odds ratios. Figure 1 shows odds ratios and their 95% credible intervals for the model parameters based on the WHO-recommended concentration of 40 µg/m$^3$ for $NO_2$.

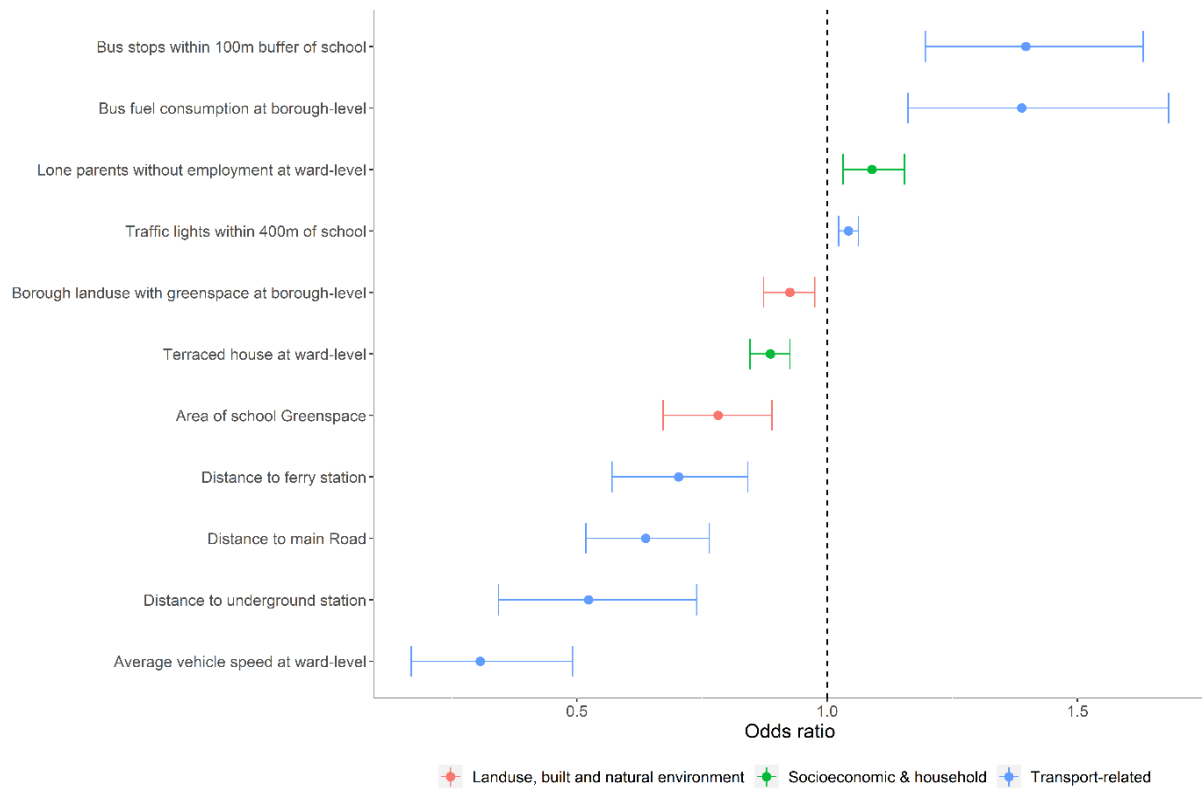

Figure 1. Odds ratios and their 95% credible intervals for the model parameters

With respect to the transport-related variables, the number of bus stops within 100m and traffic lights within 400m buffer surrounding schools were positively associated with the odds of schools exceeding the WHO-recommended concentration of $NO_2$. Specifically, one unit increase in the number of bus stops and the number of traffic lights around schools increased the odds of exceedance by 40% (CI[20%-63%]) and 4% (CI[2%-6%]), respectively. The results also showed a 39% (CI[16%-68%]) increase in the odds of exceedance for one unit (1000 tonnes of equivalent oil) increase in bus fuel consumption, reflecting the important contribution of public transport to local air quality.

We found that an increase of 5 km/h in the average road speed at ward-level decreased the odds of non-compliance by 70% (CI[51%-83%]), indicating that non-congested road networks can improve air quality around schools. Our results showed negative associations between the distance of school and the closest main road, ferry station, and underground station. Specifically, the odds of exceedance decreased by 37% (CI[24%-48%]) for every 100m increase in the distance from main roads. This is in accordance with previous research that showed air pollution concentrations decline dramatically with distance from roads [43, 44]. Similarly, a 1000m increase in the distance between school and the closest ferry station and underground station decreased the likelihood of schools exceeding the recommended concentration of $NO_2$ by 30% (CI[16%-43%]) and 48% (CI[26%-66%]), respectively.

In terms of land use, built environment and natural features, our results showed that for each unit increase in the area of school green space, the odds of non-compliance reduced by 22% (CI[11%-33%]). Similarly, one unit increase in the borough level land-use with green space decreased the odds of exceedance by 7% (CI[2%-13%]). These findings indicated the protective effects of green space against $NO_2$. Note that these natural features may not directly indicate the capacity of green space in removing pollutants (for example, through deposition on the leaves) because schools with higher area of green space and greener boroughs are generally in less polluted areas of London. However, adjusting for various factors in our model, our paper provides evidence on the protective effects of green space, perhaps functioning as a barrier between the sources of air pollutants and urban population [45].

With respect to household types, we found that schools in wards with a higher percent of terraced houses were less likely to exceed the $NO_2$ limit (odds ratio = 0.89). In terms of socioeconomic inequalities, we observed that the odds of exceedance is 9% (CI[3%-15%]) higher in schools located in wards with higher percent of lone parents with no employment.

### 4.2 Probabilities of exceeding the WHO-recommended $NO_2$ at ward and borough levels

Figure 2-a shows the spatial distribution of the ward-level estimated mean probabilities of schools' non-compliance based on the current WHO-recommended concentration of $NO_2$ (threshold value of 40 µg/m$^3$). It can be seen in Figure 2-a that the highest probability of non-compliance (more than 75% probability of exceeding) occurs in central London wards, affecting 376 schools, that is around 150,000 pupils. To investigate the sensitivity of our inferences to the pre-specified threshold value of $NO_2$, we repeated the analysis considering a hypothetical scenario in which 35 µg/m$^3$ is set as the threshold. The results, which are significantly different from Figure 2-a, are displayed in Figure 2-b. As it can be seen in Figure 2-b, a much larger area of Greater London falls into the category of high-risk wards. Specifically, with only 5 µg/m$^3$ difference between the two thresholds, a total of 1,224 schools and more than 500,000 pupils would be affected (Figure2-b).

Figure 3-a and 3-b display, respectively, the ranking of London boroughs based on their estimated mean probabilities of schools exceeding the threshold values of 40 µg/m$^3$ and 35 µg/m$^3$ and their associated uncertainties. As shown in Figure 3-a, schools located in the Inner London boroughs have in general a higher chance of non-compliance with the WHO-recommended concentration of $NO_2$. Among the Inner London boroughs, Lambeth, Hackney, Hammersmith and Fulham, Wandsworth, Lewisham, and Greenwich show a distinct trend with less than 50% probabilities of exceedance. When the threshold value was reduced to 35 µg/m$^3$, the form of the graph changes significantly (see Figure 3-b). Also, the estimated uncertainties around the higher end of the expected probabilities of exceedance become very small as most of schools result non-compliant with the hypothetical threshold. Figure 3-b implies that schools located in some Outer boroughs (Newham, Haringey, and Brent) have more than 75% chance of exceeding the threshold value of 35 µg/m³.

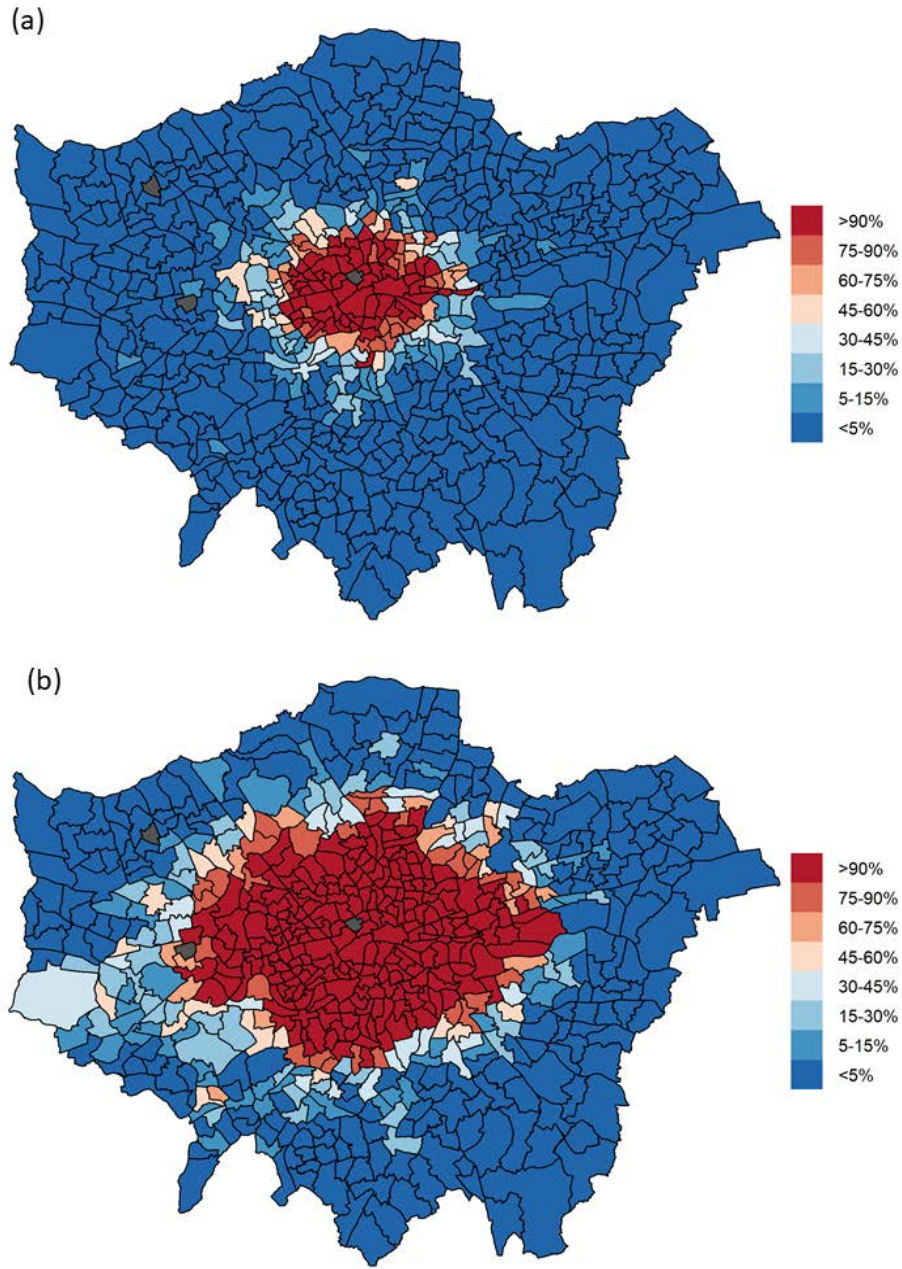

Figure 2. Mean probabilities of school non-compliance in Greater London wards: (a) based on the WHO-recommended concentration of 40 µg/m³ for $NO_2$, and (b) based on hypothetical threshold value of 35 µg/m³.

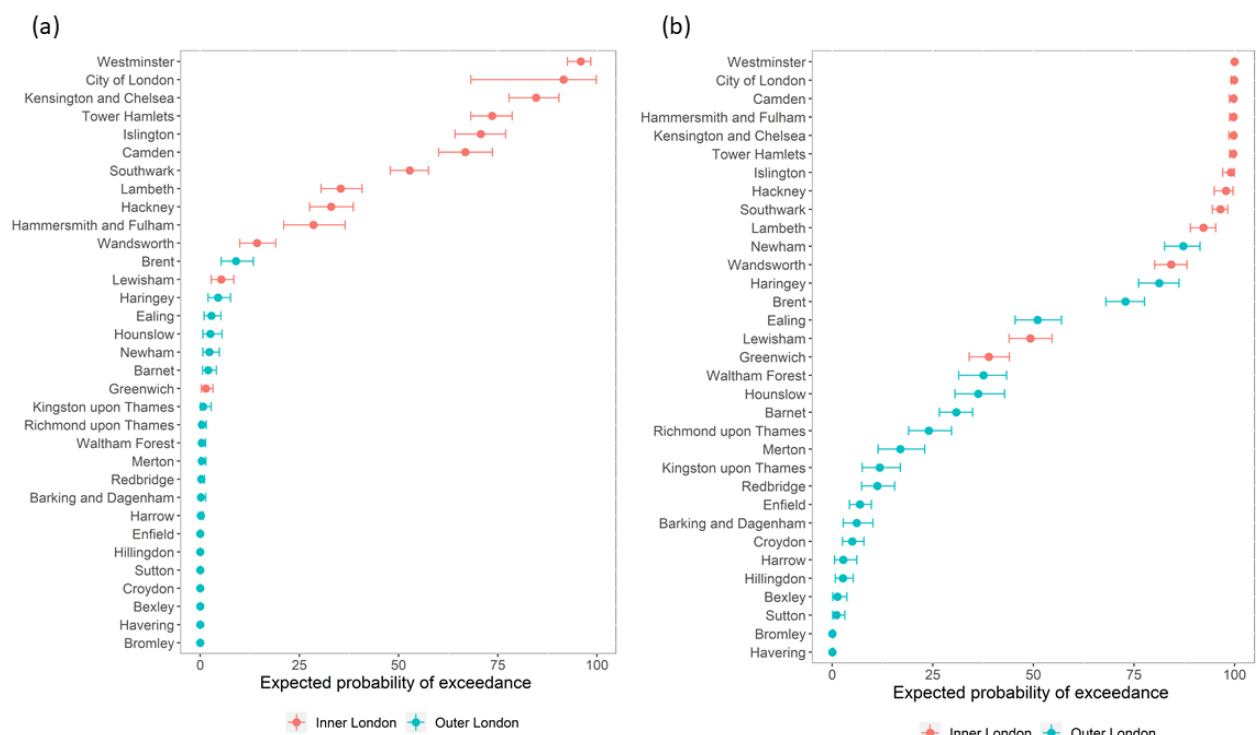

**Figure 3.** Borough ranking of mean probabilities of schools' non-compliance: (a) based on the WHO-recommended concentration of 40 µg/m$^3$ for $NO_2$, and (b) based on hypothetical threshold value of 35 µg/m$^3$.

## 5. Discussions

Our research offered a quantitative insight on how schools' attributes as well as their neighbourhood characteristics affect the odds of schools exceeding the WHO-recommended concentration of $NO_2$. Adopting a model with a multilevel structure provided additional insights on variabilities in the odds of exceedance among wards. In our research, wards are geographical units that are small enough to represent a homogeneous neighbourhood in terms of socioeconomic and environmental characteristics while being sufficiently large to have at least one school in them.

While evidence-based policies are needed to reduce emissions in London, we suggest the following strategies, some of which can be implemented by schools in partnership with local authorities, to contain unacceptable levels of air pollutants at London schools. Our results can also serve as a guideline for other urban areas. We discuss our suggested strategies as follows:

### 5.1 *Design, implement, and maintain an efficient urban transport system*

The focus here is to reduce congestion in urban settings and in particular in proximity to schools. Transportation-related features that entail frequent vehicle queuing, braking, and abrupt acceleration/deceleration often produce higher concentrations of air pollutants due to incomplete combustion and increased fuel consumption. For example, Kim et al.[46] showed that elevated pollutant concentrations around signalised junctions is a consequence of additional time in idling, acceleration, and braking. Similarly, previous research shows that an increase in the number of speed humps increased concentrations of traffic-related air pollutants [47]. Therefore, this is the reason for which we observed a strongly positive association between the number of traffic lights around schools and the likelihood of exceeding the WHO-recommended concentration of $NO_2$ at London schools.

Traffic management techniques such as optimising signal timing can create synchronized traffic lights and help maintain a smooth flow of traffic. Based on a similar logic, wards with low average operating vehicle speed (perhaps indicating traffic congestion) have higher rates of exceedance. Given that there is generally a U-shaped association between traffic emissions and vehicle speed, maintaining the operating speed at an optimal level and considering variable speed limits could be a promising solution to maintain a uniform speed and a smooth traffic flow. This in turn would result in reducing traffic emissions. A practical example of this is the implementation of a variable speed limit policy in Barcelona (Spain), where $NO_x$ emissions during peak hours were reduced by up to 17%, compared to a fixed speed limit [48].

*5.2 Reduce public transport emissions*

Our results revealed a substantial contribution of the number of bus stops around schools to the odds of exceeding the WHO-recommended concentration of $NO_2$. This calls for policies to retiring diesel buses and replacing them with those that run on a cleaner fuel, such as compressed natural gas or hybrid electric vehicles, as well as retrofitting buses with better air filtration system and devices to control pollution emissions. This should at least be of high priority in inner London wards (see Figure 2). Note that a study conducted by [37] indicates that more than 20% of $NO_2$ emissions in London are being generated by buses and coaches. Also, our finding can be partly attributed to frequent stopping/starting and idling at bus stops, which is associated with non-exhaust emissions, for example, due to tyre and break wear [49] Therefore, reducing the number of bus stops in immediate proximity to schools and reducing bus dwell time should be considered to reduce adverse effects of public transport on air quality in London schools.

*5.3 Reduce emissions at schools' doorsteps*

Localized elevated air pollution pockets around schools during arrival and dismissal times is a well-known problem [50, 51]. An intervention, supported by the Mayor of London and already spanning in 27 boroughs, has been the "Idling Action" campaign, where parents are advised to turn off car engines while waiting for their children. Another noteworthy solution would be to create dedicated drop-off zones with a limited stay time in order to reduce idling periods as well as traffic congestion around schools. Additionally, implementing carpooling programs to help parents identify families located in their neighbourhoods can reduce car trips and congestion around schools. Finally, we need to recognise busy roads as a contributory source of air pollution around schools. Currently, London has no regulation to address school siting with respect to environmental quality considerations. Urban planners, school decision-makers, and key stakeholders should work together to ensure that future schools are strategically constructed in less polluted areas of each neighbourhood, with a safe distance from high-volume roadways.

*5.4 Confine pollution at schools*

In line with previous research [52], our study provided evidence on the beneficial impact of green space at schools and surrounding areas. School board decision-makers need to commit to maintain and increase green space at schools when possible as a viable strategy to improve air quality around schools. At very least, green walls to create a barrier between air pollutants and school premises, especially those facing busy roads, should be considered.

*5.5 Reduce vehicle emissions by promoting active travel to and from school*

A study conducted by Transport for London showed that a quarter of weekday morning congestion, equivalent to 254,000 trips/day, is due to school drop-offs [53]. Initiatives that involve reduction in the use of cars would improve traffic congestion and air quality. A number of existing initiatives in London (e.g., "School Streets" and "School Superzones"), which aim to create healthy school neighbourhoods strongly support safe, active, and sustainable commute to schools. Also, the Department for Education in the UK has been facilitating active travel to schools through schemes such as various Walk to School Programmes, bikeability training, and implementing "safe streets" outside schools [54]. Schools can play a crucial role in the effective roll out of these schemes, for example, by developing organised walking/biking/scooting buses supervised by adults along a pre-specified route.

*5.6 Educational campaigns*

Education is a must and cooperation is the way to achieve it. The success of delivering the abovementioned strategies requires a collaborative community response between pupils, parents, schools, local authorities, and the Government. A part of that response involves a paradigm shift in the mind set and behaviour of individuals from childhood. This can be achieved by enhancing parents and pupils' awareness about the linkage between air quality, and health, and highlighting their responsibility in reducing air pollution through education and encouragement programmes such as school assemblies, online trainings videos, special events such as walking weeks, school activities, incentives for pupils in recognition of sustainable travel, and involvement in citizen science projects [55, 56]. The limited resources in disadvantaged schools can impact the implementation of training programmes. Local authorities and the Government should help schools in overcoming such restrictions. In addition, financial incentives are important to encourage schools to create innovative and tailored programmes with the aim of tackling the school air quality problem and improve children's health.

**Credit Author Statement**

Niloofar Shoari: Conceptualization, Methodology, Data preparation, Formal analysis, Writing, Funding acquisition. Shahram Heydari: Conceptualization, Methodology, Writing. Marta Blangiardo: Conceptualization, Methodology, Writing.

**Acknowledgements**

NS is supported by an MRC Early Career Researcher Fellowship. Infrastructure support for the Department of Epidemiology and Biostatistics was provided by the NIHR Imperial Biomedical Research Centre (BRC).

**Conflict of Interest**

All the authors of the present manuscript have no conflict of interest to declare.